\documentclass[11pt]{article}

\usepackage[OT2,OT1]{fontenc}

\usepackage{a4wide}
\usepackage{amsmath}

\usepackage{mathrsfs}
\usepackage[T1]{fontenc}
\usepackage{mathpazo}
\usepackage{setspace}
\usepackage{amsfonts}
\usepackage{amssymb}
\usepackage{amsmath}
\usepackage{epsfig}
\usepackage{latexsym}
\usepackage{color}
\usepackage{nicefrac}
\usepackage[mathscr]{euscript}
\usepackage[latin1]{inputenc}
\usepackage{cite}
\usepackage[usenames,dvipsnames]{xcolor}
\usepackage[colorlinks=true, pdfstartview=FitV, linkcolor=black, citecolor=black, urlcolor=black]{hyperref}

\numberwithin{equation}{section}

\definecolor{papercolor}{rgb}{0.25, 0 ,1}
\definecolor{markcolor}{rgb}{1, 0.2, 0.2}

\definecolor{fux}{rgb}{0.75, 0 ,1}

\author{
  \begin{minipage}{.97\linewidth}
    \vspace{1cm}
       \begin{center}
      \begin{small}
               \textbf{Robert G. Leigh},$^1$
             \textbf{Anastasios C. Petkou},$^2$ 
                     \textbf{P. Marios Petropoulos}$^3$\\
     and \\
      \textbf{Prasanta K. Tripathy}$^4$
              \end{small}
    \end{center}
    \vspace{0.5cm}
    \hspace{2.8cm}\begin{minipage}{.7\linewidth}
\begin{center}     {\it \begin{footnotesize}
\hbox{\kern-1.cm\vbox{\vskip0cm
 \begin{itemize}
               \item[$^1$] Department of Physics\\University of Illinois\\1110 W. Green Street\\ Urbana IL 61801, U.S.A.\\
{\tt rgleigh@illinois.edu}  
  \item[$^3$] Centre de Physique Th\'eorique\\ 
        Ecole Polytechnique, CNRS UMR 7644\\
        91128 Palaiseau Cedex, France\\
{\tt marios@cpht.polytechnique.fr}
\vskip0.31cm
      \end{itemize}}
\kern-5cm\vbox{
\begin{itemize}
  \item[$^2$] Department of Physics\\ 
  Institute of Theoretical Physics\\
  Aristotle University of Thessaloniki\\ 
  54124, Thessaloniki, Greece\\
      {\tt tpetkou@gmail.com}
 \item[$^4$] Department of Physics\\ 
  Indian Institute of Technology Madras\\ 
  Chennai 600 036, India\\
  {\tt prasanta@physics.iitm.ac.in}
      \end{itemize}\vskip0.05cm
}}
     \end{footnotesize}}
\end{center}
    \end{minipage}
    \vspace{0.5cm}
  \end{minipage}
}

\title{\vspace{3.cm}
 \boldmath \begin{Large}
    \textbf{The Geroch group in Einstein spaces}
  \end{Large} \unboldmath
}

\begin{document}

\begin{titlepage}
  \maketitle
  \thispagestyle{empty}

  \vspace{-15.cm}
  \begin{flushright}
    CPHT-RR051.0613\\
IITM-PH-TH-2014-1\\
  \end{flushright}

  \vspace{12cm}

  \begin{center}
    \textsc{Abstract}    
          \end{center}

Geroch's solution-generating method is extended to the case of Einstein spaces, which possess a Killing vector {{}and are thus asymptotically (locally) (anti-)de Sitter}. This includes the reduction to a three-dimensional coset space, the description of the dynamics in terms of a sigma-model and its transformation properties under the $SL(2,\mathbb{R})$ group, and the reconstruction of new four-dimensional Einstein spaces. The detailed analysis of the space of solutions is performed using the Hamilton--Jacobi method  in the instance where the three-dimensional coset space is conformal to $\mathbb{R}\times  \mathcal{S}_2$. The cosmological constant appears in this framework as a constant of motion and transforms under $SL(2,\mathbb{R})$.

\vspace{2cm} 

\end{titlepage}

\onehalfspace

\tableofcontents

\section{Introduction}

In 1970, Geroch exhibited in his seminal paper \cite{Geroch} a method for generating vacuum solutions of Einstein's equations, generalizing previous work by Ehlers \cite{Ehlers}. 
His starting point was a four-dimensional manifold $\mathcal{M}$, endowed with a metric $\mathrm{g}$ with vanishing Ricci tensor. A {{}generic, everywhere space-like or time-like,} Killing vector $\xi$ was also assumed for $\mathrm{g}$, with scalar twist $\omega$ and norm $\lambda$. A coset space $\mathcal{S}$ was further constructed  {{}as the quotient of $\mathcal{M}$ by 
the one-parameter group of motions generated by
$\xi$.}
The core of the proposed method was to set an unambiguous one-to-one mapping between  $\mathcal{S}$ and  $\mathcal{M}$, and recast the four-dimensional Einstein's equations in terms of the data on  $\mathcal{S}$: the metric $\mathrm{h}$ 
{{} on the projected space $\mathcal{S}$
orthogonal to $\xi$,}
and the scalar fields $\lambda$ and $\omega$. Any new triplet $( \mathrm{h}',  \omega',\lambda')$ satisfying that set of equations could be promoted to a new four-dimensional  vacuum solution $\mathrm{g}'$ with one isometry. 

Geroch's crucial observation was that keeping the metric $\mathrm{h}$ within the conformal class of fixed $\lambda \mathrm{h}$, new solutions could be generated as M\"obius transformations  of $\tau=\omega +i \lambda$: $ \tau'=\nicefrac{a\tau +b}{c\tau+d}$ with 
$\left(\begin{smallmatrix}      a & b \\ c & d  \end{smallmatrix}\right)$ in $SL(2,\mathbb{R})$. 
In concrete examples such as Schwarzschild--Taub--NUT solutions with mass $m$ and NUT charge $n$, the compact  
subgroup  of rotations $\left(\begin{smallmatrix}      \cos \chi & \sin \chi \\ -\sin \chi & \cos \chi  \end{smallmatrix}\right) \in SO(2)\subset SL(2,\mathbb{R})$ induced rotations of angle $2\chi$ in the parameter space $(m,n)$, while  non-compact transformations $\left(\begin{smallmatrix}      a & b \\ 0 & \nicefrac{1}{a}  \end{smallmatrix}\right)\in N \subset SL(2,\mathbb{R})$ acted homothetically,  
$(m,n)\to (\nicefrac{m}{a} , \nicefrac{n}{a} )$.

Prior to Geroch's work, important results had been obtained by Ernst \cite{Ernst:1967wx, Ernst:1968} in the slightly less general case of the Lewis--Papapetrou geometries \cite{Papapetrou:53, Papapetrou:66}, which possess two commuting Killing vectors.
In this case, solving vacuum Einstein's equations becomes a two-dimensional problem whose dynamics is governed by a two-dimensional sigma-model effective action. Further developments appeared mostly in this direction
\cite{Geroch:1972yt, Breitenlohner:1986um}, and were often oriented towards understanding the general integrability properties of the two-dimensional sigma-model \cite{belinskii, Maison:1978es, maison2, Mazur:1982}.\footnote{The literature on the integrability issues of the problem at hand is quite rich, see \cite{Bernard:2001pp, Bardoux:2013swa} for more references.
} Such analyses provide an important and  complementary perspective with respect to Geroch's algebraic solution-generating technique.

 {{}Extending the above methods for solving Einstein's equations in more general situations 
 has attracted some attention. Electrovacuum solutions were extensively studied, for example, in \cite{Alekseev:1980ew,Mazur:1983}}.{\footnote{{} For a brief summary discussing the integrability issues of vacuum 
 Einstein's  equations as well as electorvacuum equations, see \cite{Alekseev:2010mx} and references therein. 
 For recent developments on integrability issues in supergravity and string-theory solutions, see 
 \cite{Alekseev:2004zz,Alekseev:2008gh,Figueras:2009mc}.
For some recent applications of the solution-generating techniques  on asymptotically flat as well as black-ring solutions in five dimensions, see
\cite{Mishima:2005id,Iguchi:2007is}.}}
{{} However, in the presence of a cosmological constant, Geroch's approach becomes much more complicated, as we will see below, and this case is lesser investigated. Some recent examples}
in the framework of Ernst's equations can be found in \cite{Charmousis:2006fx,Caldarelli:2008pz,Astorino:2012zm}. Even though integrability properties and solution-generating techniques 
 {{}may 
 in some instances lead to already known results, they ultimately allow for a deeper understanding of the  landscape, and can} exhibit genuinely new solutions (such as the Melvin-magnetic space with cosmological constant found in \cite{Astorino:2012zm}).  {{}These perspectives and the increasing interest for (anti) de Sitter geometries} motivate our present attempt to revisit the Geroch group and integrability in the presence of a cosmological constant.

Even though Geroch insisted in starting with a vacuum solution $(\mathcal{M}, \mathrm{g})$, all the requirements necessary to translate Einstein's equations to three-dimensional terms remain valid in the more general case of Einstein spaces: $\lambda $ and $\omega$ are well-defined and together with the coset $(\mathcal{S}, \mathrm{h})$, they provide a 
 {{}complete} characterization of $(\mathcal{M}, \mathrm{g})$.  {{}In order to proceed 
 along Geroch's lines, we must be careful with the choice of conformal class for the  metric  $\mathrm{h}$.} 
 Insisting on exploring new configurations $( \omega',\lambda')$ with $\lambda' \mathrm{h}'$ held fixed to $\lambda \mathrm{h}$,  {{}as in the original work \cite{Geroch}, leads to a space of solutions that  \emph{cannot} accommodate simultaneously
 Schwarzschild-(A)dS and Taub--NUT-(A)dS spaces.} 
Overcoming this obstruction requires  the consideration of $( \mathrm{h}',  \omega',\lambda')$ with $\mathrm{h}'$ in the conformal class of $\mathrm{h}$ without further restriction. This amounts to introducing an extra scalar field $\kappa$ as the conformal factor, and investigating the dynamics of the fields $(\kappa, \omega, \lambda)$ as inherited from the four-dimensional Einstein's equations, within the conformal class of the original metric $\mathrm{h}$. 

Our results are summarized as follows. The dynamics of $(\kappa, \omega, \lambda)$ is captured by a three-dimensional sigma model with target space conformal to $\mathbb{R} \times H_2$,
{{}where $H_2$
is the 
Lobatchevski plane}. The Geroch $SL(2,\mathbb{R})$ algebra\footnote{{{}It is common in the literature to name this finite-dimensional algebra  \emph{Ehlers}, and keep \emph{Geroch} for the affine one, when present. In our study, there is no ambiguity.}} is realized as the isometry of the  target-space $H_2$-factor.  Restricting our attention to metrics $\mathrm{h}$ conformal to $\mathbb{R}\times  \mathcal{S}_2$, the problem is further reduced to particle motion on  $\mathbb{R} \times H_2$ in the presence of a scalar potential and subject to the Hamiltonian constraint. By adopting appropriately the time parameter, one reaches a Hamiltonian such that, irrespective of the value of the cosmological constant, two out of three $SL(2,\mathbb{R})$ generators are broken, although they still provide two conserved quantities when the constraint is satisfied. This leads to a unified picture, and using the Hamilton--Jacobi method integrability is explicitly demonstrated, showing in particular that the cosmological constant  {{}$\Lambda $} plays the role of a constant of motion. 
In conclusion, when $\Lambda $ is non-vanishing, Geroch's group emerges as a tool for handling the dynamics, even though only a subgroup of it provides an algebraic mapping in the space of solutions; the latter mapping transforms, among other parameters, the cosmological constant itself. 

\section{From four to three dimensions and back to four}\label{ftttf}

We assume $\mathcal{M}$ a four-dimensional manifold endowed with a Lorentzian-signature metric $\mathrm{g}=g_{ab}\text{d}x^a \text{d}x^b $, invariant under a one-parameter group of motions generated by the Killing vector $\xi$. The latter has norm and twist one-form 
\begin{eqnarray}
\lambda&=&\left\| \xi\right\|^2\\
{{}w}&=&-2 \text{i}_\xi \star \text{d}\xi
\end{eqnarray}
respectively.\footnote{We follow in this section the original presentation of Geroch (the appendix of \cite{Geroch} provides a useful complementary reading). Here ``$\star$'' is the four-dimensional Hodge duality performed with $\eta_{abcd}=\sqrt{-g}\, \epsilon_{abcd}$ ($\epsilon_{0123}=1$).
We also recall that for $\zeta$ a vector and $\varpi$ a form 
$$
\mathcal{L}_\zeta \varpi= \text{d}(\text{i}_\zeta\varpi)+ \text{i}_\zeta\text{d}\varpi,
$$
where $\mathcal{L}_\zeta$ is the Lie derivative along $\zeta$ and $\text{i}_\zeta$ is the contraction with $\zeta$. \label{magicform}} Here $\xi$ is also the Killing one-form, and obeys identically 
\begin{equation}\label{Kilprop}
\star \text{d}\star\text{d}\xi = 2  \text{i}_\xi \mathrm{Ric},
\end{equation}
where $\mathrm{Ric}$ is the four-dimensional Ricci tensor. Assuming the metric is Einstein ($\mathrm{Ric}=\Lambda \mathrm{g}$), Eq. \eqref{Kilprop} reads
\begin{equation}\label{Kilprop-Eins}
\text{d}\star \text{d}\xi = 2  \Lambda \star \xi.
\end{equation}
Using the identity 
quoted in footnote \ref{magicform}, we conclude\footnote{This conclusion actually holds whenever $\xi$ is an eigenvector of the Ricci tensor. This is trivial in vacuum and straightforward for Einstein spaces, but may occur in other instances, which we have not explored here.} that $\text{d}{{}w}
= 0$ and define the scalar twist locally as
\begin{equation}
{{}w}=\text{d}\omega.
\end{equation}

With the above data, we can define the space $\mathcal{S}$ as a quotient of $\mathcal{M}$
with respect to the action of the one-parameter group generated by $\xi$. This coset space need not be a subspace of $\mathcal{M}$ as $\xi$ may not be hypersurface-orthogonal (which would imply zero twist). There is a natural metric $\text{h}$ on $\mathcal{S}$ induced by $\text{g}$ of $\mathcal{M}$ as
\begin{equation}\label{met-h}
h_{ab}=g_{ab}-\frac{\xi_a\xi_b}{\lambda},
\end{equation}
which defines the projector onto $\mathcal{S}$ is 
\begin{equation}\label{proj-h}
h^b_a=\delta^b_a-\frac{\xi^b\xi_a}{\lambda}.
\end{equation}
For the metric \eqref{met-h}, the volume form and the fully antisymmetric tensor read:
\begin{equation}\label{vol-h}
\text{Vol}_{\text{h}} =  \frac{\mp 1}{\sqrt{\pm\lambda}}\text{i}_\xi \text{Vol}_{\text{g}}
\Leftrightarrow
 \eta_{abc}= \frac{\pm 1}{\sqrt{\pm\lambda}}\eta_{abcd}\xi^d.
\end{equation}
We will assume for concreteness $\lambda<0$ so that the Killing vector $\xi$ is time-like and $\text{h}$ is 
 {{}spatial}. This corresponds to the lower signs in \eqref{vol-h}. {{}We would like to stress, however, that the whole reduction procedure goes smoothly through when $\lambda$ is positive, \emph{i.e.} for a space-like Killing vector field.}

Following \cite{Geroch}, let us quote hereafter some basic features of the geometrical relationship between $\mathcal{M}$ and $\mathcal{S}$.
There is a natural one-to-one correspondence between tensors on $\mathcal{S}$ and tensors $T$ on $\mathcal{M}$ that satisfy $\text{i}_\xi T=0$ and $\mathcal{L}_\xi T=0$ (\emph{i.e.} transverse and invariant with respect to the Killing flow). 
Assume now a tensor $T$ on $\mathcal{S}$. It is easy to show that the covariant derivative $\mathscr{D}$ defined following this correspondence, 
\begin{equation}\label{covd-h}
\mathscr{D}_c T_{a_1\ldots a_p}^{\hphantom{a_1\ldots a_p}b_1\ldots b_q} =
h_c^\ell h_{a_1}^{m_1}\dots  h_{a_p}^{m_p}h^{b_1}_{n_1}\dots  h^{b_q}_{n_q}
\nabla_\ell T_{b_1\ldots b_p}^{\hphantom{b_1\ldots b_p}n_1\ldots n_q} 
\end{equation}
with $\nabla$ the Levi--Civita derivative on $\mathcal{M}$, coincides with the unique Levi--Civita covariant derivative on $\mathcal{S}$. This provides a Riemann tensor on $\mathcal{S}$ in terms of the Riemann tensor of $\mathcal{M}$:\footnote{Equations \eqref{R-h} are more general than Gauss--Codazzi equations, since $\xi$ needs not be hypersurface orthogonal.}
\begin{equation}\label{R-h}
\mathscr{R}_{abcd}
= h_{[a}^{\hphantom{[}p}h_{b]}^{q}
h_{[c}^{\hphantom{[}r}h_{d]}^{s}
\left(R_{pqrs}+\tfrac{2}{\lambda}\left(
\nabla_p\xi_q
\nabla_r\xi_s
+\nabla_p\xi_r
\nabla_q\xi_s
\right)\right)
\end{equation}
(the calligraphic  letters refer to $\mathcal{S}$ tensors). 

The existence of a Killing vector $\xi$ on $(\mathcal{M},\mathrm{g})$ allows us to recast the dynamics of $\mathrm{g}$ in terms of $(\mathrm{h},\omega,\lambda)$, which can all be regarded as fields on $\mathcal{S}$. For that, 
one extracts the  $\mathcal{S}$-Ricci tensor $\mathscr{R}_{ab}$ from \eqref{R-h} and further determines the  $\mathcal{S}$-Laplacians of $\lambda$ and $\omega$. The final equations are 
\begin{equation}\label{Ger-eq}
\begin{array}{rcl}
\mathscr{R}_{ab}&=&
\frac{1}{2\lambda^2}
\left(
\mathscr{D}_a \omega\mathscr{D}_b \omega-h_{ab}
\mathscr{D}^c \omega\mathscr{D}_c \omega
\right)
+\frac{1}{2\lambda}\mathscr{D}_a\mathscr{D}_b\lambda
 -\frac{1}{4\lambda^2}\mathscr{D}_a\lambda\mathscr{D}_b\lambda
 + h_a^m h_b^n R_{mn},  
\\
\mathscr{D}^2\lambda&=&\frac{1}{2\lambda}\left(
\mathscr{D}^c \lambda\mathscr{D}_c \lambda-2\mathscr{D}^c \omega\mathscr{D}_c \omega
\right)-2 R_{mn}\xi^m\xi^n,\\
\mathscr{D}^2\omega&=&\frac{3}{2\lambda}
\mathscr{D}^c \lambda\mathscr{D}_c \omega.
\end{array}
\end{equation}
Equations \eqref{Ger-eq} provide in principle new solutions $(\mathrm{h}',\omega',\lambda')$. Without ambiguity, the latter can be promoted to a new metric $\text{g}'$ with symmetry $\xi'$ on   $\mathcal{M}$. The procedure is based on the following observation: the two-form defined on $\mathcal{S}$ as\footnote{Here $\star^3_{\text{h}'}$ stands for the three-dimensional Hodge-dual with respect to $\text{h}'$.}
\begin{equation}\label{F}
\text{F}'= \frac{1}{(-\lambda')^{\nicefrac{3}{2}}}
\star^3_{\text{h}'}\text{d}\omega'
\end{equation}
is \emph{closed}. Thus, locally
\begin{equation}\label{eta}
\text{F}'=\text{d}\eta'.
\end{equation}
The field $\eta'$, \emph{a priori } defined on $\mathcal{S}$, can be promoted to a field on $\mathcal{M}$ by adding the necessary exact piece such that its normalization is 
\begin{equation}\label{eta-norm}
\text{i}_\xi \eta'=1.
\end{equation}
This defines a new Killing field on $\mathcal{M}$
\begin{equation}\label{newkil}
\xi'= \eta' \lambda' 
\end{equation}
and the new four-dimensional metric reads:
\begin{equation}\label{newmet}
g_{ab}'=h_{ab}'+\frac{\xi_a'\xi_b'}{\lambda'}.
\end{equation}

\section{The sigma-model}

Equations \eqref{Ger-eq} can be recast in a more useful manner by introducing a three-dimensional reference metric $\hat{\text{h}}$, defined as 
\begin{equation}\label{newmet3}
h_{ab}=\frac{\kappa}{\lambda}\hat h_{ab}.
\end{equation}
The dilaton-like field $\kappa$ captures one of the degrees of freedom carried by the metric $\text{h}$, and inheritates its dynamics from the latter's.
This is useful for probing mini-superspace solutions with frozen $\hat{\text{h}}$, because it allows for one gravity degree of freedom to remain dynamical, together with  
$\omega$ and $\lambda$.\footnote{The original Geroch's reference metric was defined as $\tilde h_{ab}={\lambda}h_{ab}$. Freezing $\tilde{\text{h}}$ removes all $\text{h}$ degrees of freedom, leaving only   $\omega$ and $\lambda$ as dynamical fields.}  In this instance, as advertised in the introduction, the scalar degree of freedom $\kappa$  is crucial for the system to capture e.g. mass and NUT parameters simultaneously. This issue will be further discussed in Sec. \ref{sec:HJ}. 
 
Assuming $\text{g}$ is Einstein
\begin{equation}\label{Eins-g}
R_{ab}=\Lambda g_{ab},
\end{equation}
and introducing $\tau=\omega + i \lambda$,
Eqs. \eqref{Ger-eq} read:
\begin{eqnarray}
\hat{\mathscr{R}}_{ab}&=&
-\tfrac{2}{(\tau-\bar\tau)^2}    
\hat{\mathscr{D}}_{(a}\tau\, \hat{\mathscr{D}}_{b)}\bar\tau
+\tfrac{1}{2\kappa}\left(
\hat{\mathscr{D}}_a \hat{\mathscr{D}}_b \kappa+\hat  h_{ab}
\hat{\mathscr{D}}^c \hat{\mathscr{D}}_c \kappa
\right)\nonumber\\
&&-\tfrac{1}{4\kappa^2}\left(3 \hat{\mathscr{D}}_{a}\kappa \hat{\mathscr{D}}_{b}\kappa
 +\hat  h_{ab} \hat{\mathscr{D}}^c\kappa \hat{\mathscr{D}}_c\kappa\right)
 +4i \Lambda\tfrac{\kappa}{\tau-\bar\tau}\hat h_{ab},  
\label{eq-hab}\\
\hat{\mathscr{D}}^2\tau&=&\tfrac{2}{\tau-\bar \tau}
\hat{\mathscr{D}}^c \tau\hat{\mathscr{D}}_c \tau-\tfrac{1}{2\kappa}\hat{\mathscr{D}}^c \kappa\hat{\mathscr{D}}_c \tau
-2i \Lambda\kappa,\label{eq-tau}
\end{eqnarray}
where all hatted quantities refer to the metric $\hat{\text{h}}$.
These equations describe the dynamics of the fields $(\hat{\text{h}},\kappa, \tau)$, the equation for $\kappa$ being the trace of \eqref{eq-hab}:
\begin{equation}\label{eq-kap}
\hat{\mathscr{D}}^2\kappa=
\tfrac{3}{4\kappa} \hat{\mathscr{D}}^{c}\kappa \hat{\mathscr{D}}_{c}\kappa
+
\tfrac{\kappa}{(\tau-\bar\tau)^2}    
\hat{\mathscr{D}}^{c}\tau\, \hat{\mathscr{D}}_{c}\bar\tau
-6i \Lambda\tfrac{\kappa^2}{\tau-\bar\tau}
+\tfrac{\kappa}{2}
\hat{\mathscr{R}} .
\end{equation}

Equations \eqref{eq-hab}, \eqref{eq-tau} and \eqref{eq-kap} can be obtained by extremizing, with respect to $\hat h_{ab}, \tau$ and $\kappa$, the sigma-model action  
\begin{equation}\label{gen-S}
S = \int_{\mathcal{S}} \mathrm{d}^3 x \sqrt{\hat h}\mathcal{L},
\end{equation}
with
\begin{equation}\label{gen-L-den}
\mathcal{L}=-\sqrt{-\kappa}\left(\frac{\hat{\mathscr{D}}^a\kappa
\hat {\mathscr{D}}_a\kappa}{2\kappa^2}+2
\frac{\hat {\mathscr{D}}^a\tau
\hat {\mathscr{D}}_a\bar \tau}{(\tau-\bar \tau)^2}+
\hat{\mathscr{R}}-4i \Lambda\frac{\kappa}{\tau-\bar\tau}
\right).
\end{equation}
For the  $\kappa, \omega$ and  $\lambda$ ``matter'', the target space is three-dimensional with metric read off from the kinetic term: 
\begin{equation}
\label{st-geo-n}
\mathrm{d}s^2_{\text{target}}
=\sqrt{-\kappa}\left(-
\frac{\mathrm{d}\kappa^2}
{\kappa^2}
+\frac{\mathrm{d}\omega^2+\mathrm{d}\lambda^2}{\lambda^2}\right). 
\end{equation}
This metric is conformal to $\mathbb{R}\times H_2$, which has an  $\mathbb{R}\times SL(2,\mathbb{R})$ isometry group generated by 
\begin{equation}\label{U1k}
\zeta=\frac{1}{2}\kappa\partial_\kappa,
\end{equation}
and
\begin{equation}\label{Lsl}\
\xi_+=\partial_\omega, \quad
\xi_-=\left(\lambda^2-\omega^2\right)\partial_\omega-2\omega\lambda\partial_\lambda, \quad
\xi_2=\omega\partial_\omega+\lambda\partial_\lambda,
\end{equation}
obeying 
\begin{equation}
\left[\xi_+,\xi_- \right]=-2\xi_2,\quad
\left[\xi_+,\xi_2 \right]=\xi_+,\quad
\left[\xi_2,\xi_- \right]=\xi_-.\label{Lslt}
\end{equation}
For the metric \eqref{st-geo-n}, $\zeta$ is a conformal Killing field, whereas the $\xi$s remain Killing, and generate the $SL(2,\mathbb{R})$ Geroch group.\footnote{This finite-dimensional group is sometimes called the Ehlers group, whereas its affine extension appears as Geroch.}  The quadratic Casimir of the latter is generated by the Killing tensor $\Xi= -\nicefrac{\kappa}{\lambda^2}\left(\mathrm{d}\omega^2+\mathrm{d}\lambda^2\right)$.

The  ($\kappa, \omega, \lambda$) ``matter'' potential, 
\begin{equation}
\mathcal{V}=\sqrt{-\kappa}\left(\hat{\mathscr{R}}-2 \Lambda\frac{\kappa}{\lambda}
\right),
\end{equation}
breaks part of the $SL(2,\mathbb{R})$ symmetry, which makes the integrability analysis more intricate. Let us note for the time that it is invariant under $\xi_+$ only.\footnote{The potential is also conformally invariant under $2\zeta+\xi_2$, but this observation is of no interest for the subsequent developments.} 

In the present note we will focus on a mini-superspace analysis of the integrability properties of 
\eqref{gen-S}--\eqref{gen-L-den}, leaving the general investigation for future work. We will assume that the space $\mathcal{S}$ 
is topologically $\mathbb{R}\times \mathcal{S}_2$, and the metric $\hat{\text{h}}$ of the form\footnote{{{}Had we made the choice of a space-like Killing-vector reduction, several other  options would have been possible for the -- non-positive-definite -- metric $\hat{\text{h}}$. The subsequent analysis would have been qualitatively altered, even though the principle remains unchanged}.}
\begin{equation}
\label{met-ans}
\mathrm{d}\hat{s}^2=\mathrm{d}\sigma^2+\mathrm{d}\Omega^2,
\end{equation}
where $\mathrm{d}\Omega^2$ is the two-dimensional $\sigma$-independent piece. 
We will further assume that the functions $\kappa, \omega, \lambda $ depend on the coordinate $\sigma$ only. This choice is motivated by the general structure of Einstein exact solutions such as Schwarzschild or Taub--NUT. With the ansatz \eqref{met-ans}, the Ricci tensor reads: 
\begin{equation}
\label{Ric-ans}
\hat{\mathscr{R}}_{ab} \text{d}x^a \text{d}x^b
=\frac{\hat{\mathscr{R}}}{2} \mathrm{d}\Omega^2,
\end{equation}
where $\hat{\mathscr{R}}$ is the scalar curvature of {{}$\mathrm{d}\hat{s}^2$ \emph{and} $\mathrm{d}\Omega^2$}. 

Equation \eqref{eq-hab}  has now two independent components, the trace part \eqref{eq-kap} and the transverse part. These read\footnote{The dot stands for the derivative with respect to $\sigma$.} 
\begin{equation}
\label{eq-kappa}
\ddot\kappa=\frac{\kappa}{(\tau-\bar \tau)^2}
\dot\tau\dot{\bar \tau}
+\frac{3}{4\kappa} \dot\kappa^2
+\kappa\left(\frac{\hat{\mathscr{R}}}{2} -6i \Lambda\frac{\kappa}{\tau-\bar\tau}
\right),
\end{equation}
and
\begin{equation}
\label{eq-cons}
\hat{\mathscr{R}}=
\frac{2}{(\tau-\bar \tau)^2}
\dot \tau \dot{\bar \tau}
+4i \Lambda\frac{\kappa}{\tau-\bar\tau}
+\frac{1}{2\kappa^2} \dot\kappa^2,
\end{equation}
respectively. Furthermore, Eq. \eqref{eq-tau} takes the form
\begin{equation}
\label{eqtau-n}
\ddot\tau=
\frac{2}{\tau-\bar\tau}
\dot\tau^2
-\frac{1}{2\kappa}\dot\kappa\dot\tau-2i\kappa \Lambda.
\end{equation}
Equations \eqref{eq-kappa} and \eqref{eqtau-n} describe the dynamics of $\kappa(\sigma)$ and 
$\tau(\sigma)$.  Equation \eqref{eq-cons} {{} implies on the one hand that   $\hat{\mathscr{R}}$ is a function  of $\sigma$ only, while on the other hand being the scalar curvature of $\mathrm{d}\Omega^2$, $\hat{\mathscr{R}}$ should \emph{a priori}  depend only on  $\mathcal{S}_2$ coordinates.}
It must therefore be constant and thus 
$\mathrm{d}\Omega^2$ is necessarily a metric on $S^2, E_2$ or $H_2$ with $\hat{\mathscr{R}}=2\ell$,  $\ell=1,0,-1$: 
\begin{equation}
\label{hom-ell}
\mathrm{d}\Omega^2 = \text{d}\chi^2 +f_\ell(\chi)\text{d}\psi^2, 
\end{equation}
\begin{equation}
\label{f-ell}
f_\ell(\chi)=
\begin{cases}
\sin^2\chi &\ell=1\\
\chi^2 &\ell=0\\
\sinh^2\chi &\ell=-1.
\end{cases} 
\end{equation}
Notice finally that Eq. \eqref{eq-cons} is  first-order and is thus a constraint  -- a common occurrence in solving Einstein's equations -- ensuring the consistency of the reduction \eqref{met-ans}.

Following the above Lagrangian scheme, the equations of motion  \eqref{eq-kappa} and 
\eqref{eqtau-n} are obtained using
\begin{equation}
\label{lanS2-n}
L=\frac{\sqrt{-\kappa}}{2}\left[
-\left(\frac{\dot\kappa}{\kappa}\right)^2
-4\frac{\dot\tau \dot{\bar\tau}}{(\tau-\bar \tau)^2}
-4\left(\ell-2i\Lambda\frac{\kappa}{\tau-\bar \tau}\right)\right],
\end{equation}
which is the mini-superspace version of \eqref{gen-L-den}. Equation  \eqref{eq-cons} appears now simply as the \emph{Hamiltonian constraint}
\begin{equation}
\label{H-con}
H=0.
\end{equation}
Solving Einstein's equations in the present form,  amounts to studying the motion of a particle on the three-dimensional space-time with metric \eqref{st-geo-n}, interacting with a scalar potential  
\begin{equation}\label{pot}
V=2\sqrt{-\kappa}\left(\ell- \Lambda\frac{\kappa}{\lambda}
\right),
\end{equation}
and obeying the zero-energy condition \eqref{H-con}.

For zero cosmological constant, the full Lagrangian \eqref{lanS2-n} is invariant under the full $SL(2,\mathbb{R})$
generated by \eqref{Lsl}. This allows to scan algebraically the space of solutions, even when $\kappa$ is frozen \cite{Geroch}. As it will become clear very soon, this $\kappa$ will play a genuinely dynamical role for non-vanishing $\Lambda$, where integrability is still at work despite the reduction of the symmetry due to the potential \eqref{pot}.

\section{Conservation laws and Hamilton--Jacobi method}\label{sec:HJ}

We will now proceed to discuss the role of the Geroch $SL(2,\mathbb{R})$ group as solution-generating, as well as the integrability properties of the Hamilton--Jacobi equation. For convenience, we trade the evolution parameter $\sigma$ for $\hat r$, defined as 
\begin{equation}\label{rrhat-n}
\mathrm{d} \hat r= \frac{(-\kappa)^{\nicefrac{3}{2}}}{- \lambda}\mathrm{d} \sigma,
\end{equation}
and rescale the Lagrangian to ensure that the action remains invariant:  $\int \text{d}\sigma L = \int \text{d}\hat r \hat L$. The reader may wonder why such a transformation is performed. In \eqref{lanS2-n} the
$SL(2,\mathbb{R})$ symmetry is reduced by the $\Lambda$ term of the potential \eqref{pot}. The integrability properties of the model, however, are not affected as long as the total number of commuting charges remains unaltered. A unified treatment is therefore possible since non-vanishing $\Lambda$ destroys only non-commuting symmetries. Transformation \eqref{rrhat-n} is designed to alter the symmetry, and render its residual part independent of $\Lambda$. Indeed, using the generalized momenta
\begin{equation}
p_\kappa=\frac{1}{\lambda}\frac{\text{d}\kappa}{\text{d}\hat r} ,\quad
p_\omega=-\frac{\kappa^2}{\lambda^3}\frac{\text{d}\omega}{\text{d}\hat r},\quad
p_\lambda=-\frac{\kappa^2}{\lambda^3}\frac{\text{d}\lambda}{\text{d}\hat r},
\end{equation}
we move to the Hamiltonian formalism with 
\begin{equation}\label{newrescaled-n}
\hat H = \frac{\lambda}{2} p_\kappa^2 - \frac{\lambda^3}{2\kappa^2} (p_\omega^2 + p_\lambda^2)
+2\ell \frac{\lambda}{\kappa} - 2 \Lambda.
\end{equation}
As advertised, the symmetries and integrability properties of this Hamiltonian do not depend on $\Lambda$. The latter appears as a constant potential, and due to the conservation of energy, acquires the status of  a simple constant of motion. 

In the Hamiltonian formalism, the Geroch algebra \eqref{Lsl}--\eqref{Lslt} is realized in terms of Poisson brackets with the following phase-space functions:
\begin{eqnarray}
\hat F_+ &=& p_\omega, \label{fhp}\\
\hat F_- &=& - 2 \omega \lambda p_\lambda - (\omega^2 - \lambda^2) p_\omega - 4 \Lambda \omega \hat r,  \label{fhm}\\
\hat F_2 &=& \omega p_\omega + \lambda p_\lambda + 2 \Lambda\ {\hat{r}}.  \label{fh2}
\end{eqnarray}
It should be stressed here that the extra $\hat r$-dependent terms in $\hat F_+$ and $\hat F_2$ are not necessary for reproducing the $SL(2,\mathbb{R})$ algebra.\footnote{It would have been enough to consider the functions $\xi_i^\mu p_\mu$, $i=\pm,2$.} They are needed, however, for $\hat F_2$ to generate a first integral. Indeed, the Poisson brackets with the Hamiltonian\footnote{We use $\left\{F,G \right\}=\sum_\mu \left(\frac{\partial F}{\partial p_\mu}\frac{\partial G}{\partial q^\mu}
-\frac{\partial G}{\partial p_\mu}\frac{\partial F}{\partial q^\mu}\right)
$ and remind that $
\frac{\mathrm{d}F}{\mathrm{d} \hat r}=\frac{\partial F}{\partial \hat r} + \{H,F\} 
$.}  read:
\begin{eqnarray}
\big\{\hat H,\hat F_+\big\} &=&  0 ,  \label{Hp}\\
\big\{\hat H,\hat F_2\big\} &=& - \hat H - 2 \Lambda  ,  \label{H2}\\
\big\{\hat H ,\hat F_-\big\} &=& 2 \omega \hat H + 4 \Lambda \Big(\omega + \frac{\hat r \lambda^3 p_\omega}{\kappa^2}\Big), \label{Hm}
\end{eqnarray}
whereas
\begin{eqnarray}
\frac{\mathrm{d}\hat F_+}{\mathrm{d}\hat r} &=& 0, \label{dfp}\\
\frac{\mathrm{d}\hat F_2}{\mathrm{d}\hat r} &=& - \hat H,  \label{df2}\\
\frac{\mathrm{d}\hat F_-}{\mathrm{d}\hat r} &=& 2 \omega \hat H + 4 \Lambda  \frac{\hat r \lambda^3 p_\omega}{\kappa^2}. \label{dfm}
\end{eqnarray}

The above analysis shows that on the $\hat H =0 $ surface, $\hat F_+$ and $\hat F_2$ are conserved.  This reflects the solution-generating role of the Geroch subgroup $N\subset SL(2,\mathbb{R})$. 
For the case $\Lambda = 0$, all three dynamical functions $F_{\pm}$ and $F_2$ are conserved. Nevertheless, the integrability properties remain unaltered despite the reduction of the $SL(2,\mathbb{R})$ first integrals.
This is not surprising since only the commuting first integrals are relevant for the integrability issues. 
We will here discuss these properties using the Hamilton--Jacobi method. This has the virtue 
to work irrespective of the value of $\Lambda$ because the latter will appear as a simple integration constant. 

The equation to solve is
\begin{equation}\label{HJ-n} 
\hat H \left(\frac{\partial S}{\partial q^i},q^i
\right)
+\frac{\partial S}{\partial \hat r}
=0.
\end{equation}
The aim is to find the principal solution $S\left(q^i,\hat r; \alpha_i
\right)$  with 3 arbitrary constants $\alpha_i$. From this solution, the equations of motion become algebraic,  $\beta^i=\frac{\partial S}{\partial \alpha_i}$
with $\beta^i$ new arbitrary constants and momenta  $p_i=\frac{\partial S}{\partial q^i}$.

In the case at hand, the separation of variables in \eqref{HJ-n} is partial. Two commuting first integrals can be used for that purpose:  $\hat F_+$ and $\hat H$ with values $2\nu $ and $\hat E$. Hence
\begin{equation}
\label{prin1-n}
S= W + 2\nu  \omega - \hat E \hat r,
\end{equation}
where $W$
satisfies 
\begin{equation}\label{HJW-n} 
\frac{\lambda}{2} \left(\frac{\partial W}{\partial \kappa}
\right)^2
-
\frac{\lambda^3}{2\kappa^2} \left( \left(\frac{\partial W}{\partial \lambda}
\right)^2
+
4\nu ^2
\right)
+
2\ell \frac{\lambda}{\kappa}
=
 \hat E+2\Lambda\equiv E.
\end{equation}
A solution $W(\kappa, \lambda; E,\nu ,\alpha)$, depending on an extra constant $\alpha $ must be found for this equation.  
Once the Hamilton--Jacobi procedure is completed, $E$ will have to be set equal to $2\Lambda$.

With or without cosmological constant, Eq. \eqref{HJW-n} is not further separable. It is however integrable and, after a few technical steps, we find:
\begin{equation}
\label{prin2-n}
W= \frac{1}{6\alpha^2 } 
\sqrt{\frac{2\alpha\kappa}{\lambda} - \nu ^2}
\Big( E \Big(\nu ^2 + \frac{\alpha\kappa}{\lambda} \Big) 
+ 6 \alpha \left( 2\alpha \lambda - \ell \right)\Big).
\end{equation} 
%
%
The general solutions can now be obtained upon substituting the above expression in Eq. (\ref{prin1-n}) and differentiating 
with respect to $\alpha_1=E, \alpha_2=\nu $ and $\alpha_3=\alpha$. Let us introduce
a new set of constants
\begin{equation}\label{ms} 
\beta^1=\frac{\partial S}{\partial E}, \quad \beta^2=\frac{\partial S}{\partial \nu } , 
\quad \beta^3 =\frac{\partial S}{\partial \alpha} .
\end{equation} 
A straightforward computation leads to:
\begin{eqnarray}
\beta^1 &= & \frac{1}{6\alpha^2}\sqrt{\frac{2\alpha\kappa}{\lambda} - \nu ^2}
\left( \frac{\alpha\kappa}{\lambda} +  \nu ^2 \right) -  \hat r,  \label{bet1}\\
\beta^2 \alpha^2 \sqrt{\frac{2\alpha \kappa} { \lambda} -  \nu ^2}
&=& \frac{E}{2}   \nu  \left( \frac{\alpha \kappa}{\lambda} -  \nu ^2\right) 
+ \alpha  \left(\ell \nu   - 2 \alpha \nu  \lambda + 2 \alpha \omega \sqrt{\frac{2\alpha\kappa}{\lambda} - \nu ^2}\right), \label{bet2}
\\
3 \alpha^3 \beta^3  \sqrt{\frac{2\alpha \kappa} { \lambda} -  \nu ^2}
&=& \frac{\kappa}{\lambda}\left(
\frac{E}{2} \bigg(2 \nu ^4 \frac{\lambda}{\kappa} - 2 \alpha \nu ^2  - \alpha^2\frac{\kappa}{\lambda} \right) 
+ 3 \alpha  \bigg(2\alpha^2 \lambda + \ell \alpha  - \ell \nu ^2 \frac{\lambda}{\kappa}
\bigg)\bigg).\label{bet3}
\end{eqnarray}

The above three algebraic equations can be solved to obtain the general expressions for $\kappa,\omega,\lambda$, as functions of $\hat r$ and of the set of constants of motion $E,\nu,\alpha,\beta^1, \beta^2$ and $ \beta^3$. Not all $\beta$s are relevant, though: by a shift of $\hat r$ we can absorb $\beta^1$, while  $\beta^2$ only translates  $\omega$ without altering  $\Omega$. We will set them to zero without loss  
of generality and keep only  $\beta^3$ renamed as 
\begin{equation}
\beta^3=\sqrt{2}\frac{m}{\alpha^{\nicefrac{3}{2}}}.
\end{equation} 
We now proceed to solving Eqs. \eqref{bet1}--\eqref{bet3}. The first is actually an equation for 
$\nicefrac{\kappa}{\lambda}$:
\begin{eqnarray}
\left(36 \hat r^2 \alpha^4 + \nu ^6\right) - 3 \alpha^2 \nu ^2 \left(\frac{\kappa}{\lambda}\right)^2 
- 2 \alpha^3  \left(\frac{\kappa}{\lambda}\right)^3 = 0.
\end{eqnarray}
This is a cubic polynomial in $\nicefrac{\kappa}{\lambda}$ with negative discriminant and hence admits a 
unique solution for  $\nicefrac{\kappa}{\lambda}$ as a function of $\hat r$. It can be noticed that this equation
is factorized upon the substitution
\begin{equation} \label{rhattor}
\hat r = \frac{1}{\sqrt{2\alpha}} \left(\frac{r^3}{3} + r n^2\right),
\end{equation}
with $\nu$ traded for $n$ as
\begin{equation} \label{NUTon}
n=\frac{\nu}{\sqrt{2\alpha}}.
\end{equation}
We find 
\begin{equation} 
\frac{\kappa}{\lambda} =\left(r^2 + n^2\right).
\end{equation}
Substituting the resulting value in Eq. (\ref{bet3}) and setting $E=2\Lambda$, we find a linear equation in $\lambda$ which can be 
straightforwardly solved as a function of $r$:
\begin{equation}
\label{sol-lam}
\lambda = -\frac{\Delta}{2 \alpha(r^2 + n^2)},
\end{equation}
where
\begin{equation}
\Delta = \ell (r^2 - n^2)  - 2 m r
- \frac{\Lambda}{3} \left(r^4 + 6 r^2 n^2 - 3 n^4\right).
\end{equation}
Hence
\begin{equation}
\label{sol-kap}
\kappa = -\frac{\Delta}{2\alpha}. 
\end{equation}
Solving for $\omega$ from Eq. (\ref{bet2}), we find: 
\begin{equation}
\label{sol-ome}
\omega = -\frac{n}{3\alpha}\left(\Lambda r+\frac{3\ell r-3m-4\Lambda n^2 r}{r^2+n^2} \right).
\end{equation}
 
It is clear from the above equations that the arbitrary parameter $\alpha$ plays a normalization role.  
The interpretation of the general solution at hand \eqref{sol-lam}--\eqref{sol-ome} is straightforward upon setting $\alpha$ to 
\begin{equation}
\alpha = \frac{m^2+\ell^2 n^2}{2}.
\end{equation} 
The four-dimensional Einstein metric reconstructed following the steps of Sec. \ref{ftttf} is a three-parameter family, where $(m,n)$ are the mass and NUT charge,\footnote{From the Hamiltonian perspective of the mini-superspace analysis, $m$ appears as a ``coordinate'', while $n$ is a ``momentum'', as expected for the mass and NUT charge. Actually, for vanishing cosmological constant, the $SL(2,\mathbb{R})$ automorphism exchanging $\hat F_+$ and $\hat F_-$ acts on the Hamiltonian system as a coordinate--momentum duality, and permutes $m$ and $n$ in the solution. For non-zero $\Lambda$, this automorphism breaks down together with the $(m,n)$ duality map.} and $\Lambda$ is the cosmological constant. It reads:
\begin{equation}
\text{g}= -\frac{\Delta}{(m^2+ \ell^2 n^2)(r^2+n^2)}\left(\text{d}T+4n\sqrt{m^2+\ell^2 n^2}f_\ell(\nicefrac{\chi}{2})\text{d}\psi\right)^2 + (r^2+n^2)\left(\frac{\text{d}r^2}{\Delta}+\text{d}\Omega^2\right)
\end{equation} 
with $\text{d}\Omega^2$ and $f_\ell(\chi)$ given in \eqref{hom-ell} and \eqref{f-ell}.

A few remarks are in order at this stage. The first concerns the role of $\kappa$, advertised as being central  for the mini-superspace analysis of \eqref{gen-L-den} to capture both Schwarzschild and Taub--NUT (A)dS spaces. In Geroch's original method, \emph{i.e.} for zero cosmological constant, 
the M\"obius transformation $\tau\to \nicefrac{a\tau +b}{c\tau+d}$, which induces a motion in the $(m,n)$ parameter space, leaves $\lambda\text{h}$ invariant. Indeed, the , $m$- and $n$-dependence of 
the metric $\lambda\text{h}$ can be reabsorbed in the radial-coordinate redefinition (only $\ell=1$ is relevant for vacuum solutions)
\begin{equation}
\cosh \sigma=\frac{r-m}{\sqrt{m^2+n^2}},
\end{equation} 
leading to $\kappa = -\sinh^2\sigma$ and 
\begin{equation}
\label{lamh}
-\lambda\text{h}=-\kappa\hat{\text{h}}=
\sinh^2\sigma
\left(\text{d}\sigma^2+\text{d}\Omega^2\right)
.
\end{equation} 
Hence $\kappa$ is constant over the parameter space, and re-expressing $\lambda\text{h}$ as $\kappa \hat{\text{h}}$ does not enlarge the solution space. For non-vanishing $\Lambda$, however, neither $m$ nor $n$ can be eliminated from $\kappa$ (or from $\lambda\text{h}$) by coordinate redefinition:  $\kappa$ varies inside the $(m,n)$ plane. Had we dismissed $\kappa$, it would have been impossible to explore the whole space of solutions with the mini-superspace ansatz \eqref{met-ans} -- \emph{i.e.} with the original Geroch's ansatz of fixed $\tilde{\text{h}}=\lambda\text{h}$.

The second remark is about Geroch's $SL(2,\mathbb{R})$. For vanishing $\Lambda$, the whole group leaves the Lagrangian density
\eqref{gen-L-den} invariant and acts algebraically on $(m,n)$. In the case at hand, Eqs. \eqref{Hp}, \eqref{H2} and   \eqref{dfp}, \eqref{df2} show that only $\hat F_+$ and  $\hat F_2$, generating the subgroup $N\subset SL(2,\mathbb{R})$, survive the cosmological constant. Furthermore, owing to 
\eqref{H2}, the algebraic action of $N$ alters not only $m$ and $n$ but also $\Lambda$. This is consistent with the fact that in the present formalism,  $\Lambda$ appears as a constant of motion. Under transformations in $N$, $\tau\to a(a\tau+b)$, the parameters get modified as $m\to \nicefrac{m}{a}, n\to \nicefrac{n}{a}$ and $\Lambda \to a^2 \Lambda$.\footnote{Notice for completeness that $\alpha\to \nicefrac{\alpha}{a^2}$, whereas the constants $\beta^i$ transform as $\beta^1\rightarrow\nicefrac{\beta^1}{a^3}, \beta^2\to  a^2 \beta^2 + 2ab$ and $\beta^3\rightarrow a^2\beta^3$, respectively. The parameter $b$ 
plays no substantial role: it formally enables us to set $\beta^2$ to zero. }

Finally, let us remind that in the absence of a cosmological constant, $\hat F_+, \hat F_-$ and $\hat F_2$ transform in the adjoint representation of $SL(2,\mathbb{R})$, with  $\hat F_+ \hat F_- +\hat F_2^2$ the quadratic Casimir \cite{Geroch}. One may wonder how much of this property survives $\Lambda$.
Here $\hat F_-$ is not conserved, but thanks to \eqref{dfm}, we obtain using the explicit solution (up to an arbitrary constant, set here to zero):
\begin{equation}
\hat F_-  -4 \Lambda \int\text{d}\hat r \frac{\hat r \lambda^3 p_\omega}{\kappa^2} 
= \frac{2n\ell^2 }{\left(m^2+\ell^2 n^2\right)^{\nicefrac{3}{2}}}+ \frac{2n^3\Lambda\left(6\ell  - 23 n^2\Lambda\right)}{9 \left(m^2 + \ell^2 n^2\right)^{\nicefrac{3}{2}}}.
\end{equation}
This, together with $\hat F_+ = 2n\sqrt{m^2+\ell^2 n^2}$ and $\hat F_2 = \nicefrac{2m}{\sqrt{m^2+ \ell^2 n^2}}$ gives the would-be quadratic Casimir
\begin{eqnarray}
\hat F_+\left(\hat F_--4 \Lambda \int\text{d}\hat r \frac{\hat r \lambda^3 p_\omega}{\kappa^2}\right)+\hat F_2^2
= 4 \left( 1 + \frac{n^4\Lambda\left(6\ell  - 23 n^2\Lambda\right)}{9  \left(m^2 + \ell^2 n^2\right)}\right),
\end{eqnarray}
which is conserved, and invariant under the action of the subgroup $N$ only. For zero $\Lambda$ we recover the already quoted result.

\section{Outlook}

Summarizing our findings, we would like to stress two achievements of the present work.

Firstly, we exhibited the realization of the $SL(2,\mathbb{R})$ group in the dynamics of the three-dimensional coset space, obtained by reducing an Einstein space along some Killing field. This dynamics is captured by a sigma model, whose three-dimensional target space is non-compact and has $SL(2,\mathbb{R})$ isometry. Part of this symmetry is broken down to a subgroup of $SL(2,\mathbb{R})$ by the potential term. Only this subgroup generates algebraically new solutions \emph{\`a la} Geroch.

Secondly, we studied the integrability properties of a mini-superspace reduction, using the Hamilton--Jacobi method. Irrespective of the value of the cosmological constant, the Hamilton--Jacobi equation is partially separable but fully integrable. One separation constant is the energy and is forced to be equal to the cosmological constant. Had we performed our analysis starting with Ricci-flat four-dimensional space--time, the possibility of finding new solutions of the general Einstein type would have arisen automatically. This option appears here because we have kept dynamical the dilaton-like field $\kappa$, and chosen appropriately the coordinate $\sigma$ (Eq. \eqref{rrhat-n}) -- at the expense of breaking the $SL(2,\mathbb{R})$ even at $\Lambda = 0$. Mass, NUT charge and cosmological constant appear as the non-trivial first integrals, \emph{i.e}, the ones that cannot be reabsorbed in field redefinitions. All these parameters transform under the algebraic action of the unbroken symmetry group $N$.   The compact $SO(2)\subset SL(2,\mathbb{R})$ subgroup plays no role in this approach and merely appears as an accidental symmetry at $\Lambda = 0$, bringing no new constant of motion.

This status of the cosmological constant is an important observation, to which we plan to return in the study of the general sigma model. The latter includes the axially symmetric case first introduced by Ernst \cite{Ernst:1967wx, Ernst:1968}, and  captures, {\em e.g.} the Kerr solution, which has (A)dS extensions. In this instance, \emph{i.e.}, with two commuting isometries, the corresponding two-dimensional model can be analyzed by standard Lax-pair and inverse-scattering techniques, presently under investigation.

\section*{Acknowledgements}
{\small  The authors benefited from discussions with D. Bernard, G.~Bossard, M.~Caldarelli, C.~Charmousis, A. Hanany, S. Katmadas, A. Lakshminarayan, A. Mukhopadhyay, H. Nicolai, K.~Sfetsos, K. Siampos, K. Skenderis, A. Virmani and R.~Zegers. They also thank their institutes for hospitality and financial support. Rob Leigh and Marios Petropoulos thank the BIRS center and the organizers of the 2013 workshop \textsl{Holography and Applied String Theory}, where the ideas developed here where exchanged.  The feedback from Southampton University group and IHES was also valuable. The work of A. C. Petkou was supported in part by European Union's Seventh Framework Programme
under grant agreements (FP7-REGPOT-2012-2013-1) no 316165, the EU program "Thales" MIS 375734
and was also cofinanced by the European Union (European Social Fund, ESF) and Greek national funds through
the Operational Program Education and Lifelong Learning" of the National Strategic
Reference Framework (NSRF) under Funding of proposals that have received
a positive evaluation in the 3rd and 4th Call of ERC Grant Schemes. This research was supported by the ANR contract  05-BLAN-NT09-573739, the ERC Advanced Grant  226371, the ITN programme PITN-GA-2009-237920, the IFCPAR CEFIPRA programme 4104-2, the U.S. Department of Energy contract FG02-91-ER4070, the EU contract FP7-REGPOT-2008-1: Crete HEPCosmo-228644 and the grant \textsl{ARISTEIA II: 90748} from the Greek Secretariat of Research and Technology.}

\end{document}